\documentstyle[twocolumn,prc,aps]{revtex}
\input epsf

\def\lamds{\Lambda_{DS}}
\def\lamnda{\Lambda_{NDA}}
\def\drbar{\overline{DR}}

\begin{document}

\twocolumn[\hsize\textwidth\columnwidth\hsize\csname @twocolumnfalse\endcsname
\hfill hep-ph/9803258\\
\hfill  CERN-TH/98-63\\
               \hfill   MIT-CTP-2720

\title{Implication of Exact SUSY Gauge Couplings for QCD
}
\author{L.Randall$^{1,2}$, R. Rattazzi$^2$, E.\ Shuryak$^{1,3}$,  }
\address{
$^1$ MIT, Cambridge MA \\
$^2$ CERN, Geneva, Switzerland \\
$^3$ Department of  Physics and Astronomy,
 State University of New York at Stony Brook,
 NY 11794-3800}

\date{\today}
\maketitle  

\begin{abstract}
The phase structure of SUSY gauge theories can be very different
from their nonsupersymmetric counterparts. Nonetheless,
there is interesting information which might be gleaned
from detailed investigation of these theories. In particular,
we study the precise meaning of the strong interaction scale
$\Lambda$. We ask whether one can meaningfully apply
naive dimensional analysis and also ask whether
 the study of supersymmetric
theories can shed light on the 
 apparent discrepancy between the perturbative scale $\Lambda_{QCD}$
 and the ``chiral lagrangian'' scale $\Lambda_\chi$. 
We show that in $N=1$ supersymmetric
Yang Mills theory, ``naive dimensional analysis'' seems to work well,
with $\Lambda_\chi$ consistently equal to the scale
 at which the perturbatively evolved  physical coupling becomes of order
$4 \pi$.  We turn to $N=2$ theories to understand better the effect
of instantons in accounting for the QCD discrepancy between scales.
In
$N=2$ supersymmetric $SU(2)$ the instanton
corrections are known to all orders from the Seiberg-Witten
solution and give rise to a finite scale ratio
between the scale at which the perturbatively evolved and ``nonperturbatively
evolved'' couplings blow up. Correspondingly, instanton effects
 are important even when the associated
perturbatively evolved gauge coupling only gives $\alpha$ of order 1 (rather
than $4 \pi$).  
We compare the $N=2$ result to
 instanton-induced corrections in QCD, evaluated using lattice data and
the instanton liquid model, and find a remarkably similar
behavior.

 \end{abstract}

\vspace{0.1in}
]

\begin{narrowtext}
\newpage

\section{Introduction}
 
One usually employs the notion of a ``chiral
lagrangian" or sigma model to describe the low-energy
degrees of freedom of a strongly interacting gauge theory
below a scale which we will refer to as $\Lambda_{NDA}$,
where $\Lambda_{NDA}$  sets the scale for the cutoff
and the suppression of higher derivative interactions,
with coefficients determined by ``Naive Dimensional Analysis" \cite{gm}.
Although there is a potential ambiguity in its
precise definition, one would expect  $\Lambda_{NDA}$ 
to be the point where  the
original perturbative formulation 
of the field theory becomes impossible, so that  perturbation
theory requires the use of  effective 
degrees of freedom and effective Lagrangians. The
well-studied example of this phenomenon is QCD,
where the cutoff scale for the chiral Lagrangian is of order 1 GeV.

It can be argued that the $\Lambda_{NDA}$ scale should
be the scale at which perturbation theory fails completely
in the sense that a coupling is of order $4 \pi$ so
that a loop expansion is no longer possible \cite{nda}.  However,
in QCD this higher scale is mysterious, as the running QCD coupling
should be of order $\alpha \sim 4\pi$ at the $\Lambda_{NDA}$ scale of
order $1 {\rm GeV}$. According to the perturbative evolution,
this is manifestly not the case. Related to this puzzle
is the question of why the $1 {\rm GeV}$ scale is relevant at all,
in light of the fact that 
  both the scales
at which the coupling blows up  ($\Lambda_{pert}$) and the
confinement scales  ($\Lambda_{conf}$) are of order   250 MeV.
  Although
one might take the position that order of magnitude estimates
might not lead to a better understanding, the whole point of
naive dimensional analysis is to include as well as possible any
large factors (of order $4 \pi$) which can be readily identified.
We will discuss some exact results in supersymmetric theories
to see whether they give
some insight into this discrepancy between scales.
This might seem absurd in that the phase structure of supersymmetric
theories can differ  qualitatively from  a
nonsupersymmetric   theory. 
However, there are some questions
which one has about QCD that can be meaningfully
asked about  solvable
supersymmetric theories.  These include
the question of whether or not there can exist two distinct
meaningful physical scales, whether or not the coupling
becomes nonperturbative at one or the other of these scales,
and which nonperturbative effects are significant at any given scale.

We will show that in $N=1$ SYM theory  the scale at which the perturbative
expansion breaks down according to the all order beta function
is $\sim (8 \pi^2N)^{1/3}$ bigger than the scale at which the
holomorphic  coupling blows up. We will show that
the larger scale is in qualitative agreement
with the NDA scale as determined from the exact 
 gluino condensate.   The basic conclusion is
that there is one scale which determines the physics.
This scale is the NDA scale. 
  However, the fact that
the perturbatively evaluated coupling blows up at the NDA scale
is not the same as  the 
behavior of   QCD, where the NDA scale is not associated with the
perturbative blow up of the coupling.
 It does however suggest that NDA is a reliable tool 
that can be reasonably applied to estimate Kahler potential
terms in phenomenological applications of strongly coupled
supersymmetric theories.

We then argue that the mismatch of scales in QCD
could be  due to nonperturbative corrections.   That these effects
can be large even when the perturbative coupling
is relatively small will be shown using the exact
results from $N=2$ SU(2) in Section 3. 
  In Section 4   we
will  veer from exact results back to QCD.
We use the instanton-induce corrections to the effective charge
defined by  Callan, Dashen and Gross (CDG) \cite{cdg}, which includes
integrating out instantons of sizes less than some $
\rho_{max}$. Those integrals are done using
 available lattice data for SU(2), SU(3) pure gauge theories as well as
the instanton liquid model (fitted to QCD). 
The results display  remarkable similarity to behaviour of the N=2 theory,
which indicate the   potential significance of instanton
effects.

\section{ $N=1$ Supersymmetric Yang-Mills}

In this section, we will consider $N=1$ SYM for which
the exact $\beta$ function is known.  We will demonstrate
that the exact coupling reproduces the assumptions
underlying NDA. In particular, we will show
that two relevant scales can be defined, the
first at which the holomorphic coupling blows up  and the
second higher scale at which the true physical coupling
becomes non-perturbative. The crucial distinction \cite{sv}
is between the holomorphic (or ``Wilsonian'') coupling, which 
runs only at one-loop,
and the ``physical'' 1PI coupling which receives corrections
at all orders in perturbation theory. We will argue
that the true scale of the strong interactions is the latter. 
 
We first define the different scales. The quantity
which is generally used to construct the holomorphic superpotential
is defined by
\begin{equation}
\Lambda_{DS}=\mu e^{-8 \pi^2/b_0 g_h^2(\mu)}
\label{lambdads}
\end{equation}
where $g_h^2$ is the Wilsonian coupling constant, defined as the coefficient
 of the gauge kinetic operator in the lagrangian when written
in a manifestly holomorphic form.  Notice that
although $g_h$ is scheme dependent, $\Lambda_{DS}$ is not. This is because the
overall coefficient in eq. \ref{lambdads}, which we take to be $1$ in $\drbar$
\cite{fp}, also changes accordingly.

Another useful scale to define is $\Lambda_\infty$,
which we define as the scale at which the physical coupling
becomes non-perturbative  according to the exact NSVZ $\beta$-function
\begin{equation}
\Lambda_\infty=\mu \left( {8 \pi^2 \over g^2(\mu) N }\right )^{1/3}
 e^{-8 \pi^2/
b_0 g^2(\mu)} = \left( {8 \pi^2 \over N}\right)^{1/3} \Lambda_{DS}
\label{linf}
\end{equation}
where we have used the relation between the holomorphic
and physical 1PI coupling \cite{nsvz} in the last equality.
Notice that according to  NSVZ evolution the coupling constant never blows up.
Instead it reaches a maximum at $\mu\simeq(1.39)\Lambda_\infty$
where $Ng^2/8 \pi^2=1$ and where perturbation theory breaks down. Notice also 
that the scale $\Lambda_{2-loop}$, defined by truncating the RG beyond 
two loops (where it is anyway scheme dependent) differs from $\Lambda_{\infty}$
only by a factor  $(1+{g^2 N /8 \pi^2})^{1/3}$. 

This relation between the Wilsonian and 1PI couplings
was originally obtained by NSVZ \cite{nsvz} by considering
the $SU(N)$ SYM instanton amplitude, which is proportional to
\begin{equation}
A_{inst}=\mu^{3N} {e^{-8 \pi^2/g^2(\mu)} \over g^{2N}(\mu)} \equiv
\mu^{3N} e^{-8\pi^2 \over g_h^2(\mu)} =\Lambda_{DS}^{3N}.
\end{equation}
The relation between $g_h^2$ and $g^2$ is also simply obtained 
by considering the
nontrivial Jacobian which occurs when going from
holomorphic to canonical gauge fields \cite{arkani}. The Jacobian can be
evaluated by using the Konishi anomaly together with the
known fact that the beta function vanishes beyond 1-loop
in $N=2$ SYM.

The important thing to notice in the two scales we have defined is that
they differ by a  reasonable factor, namely $(8 \pi^2/N)^{1/3}$,
which is about 3 for low $N$.

Now we turn to Naive Dimensional Analysis, which was
applied to supersymmetric theories in Ref. \cite{nda} and
applied in various model-building studies 
\cite{kaplan,cohen,nelstra,plateau,infla,luty} to estimate
non-perturbative contributions to  the Kahler potential. 
According to NDA, operators composed of fields
which are noncanonically normalized (with a coefficient
of their kinetic term $1/g^2$) are expected
to have expectation values which scale according to the
power of $\Lambda_{NDA}$ given by the dimension of the
operator, with no additional factors of $4 \pi$.  This
can be derived by requiring that all orders in the loop expansion
(with a cutoff) give comparable contributions to operators \cite{gm}
or   directly by rescaling fields in a suitably defined Wilson 
effective action \cite{nda}. Notice that 
in a supersymmetric theory the loop expansion is not a power
series in the holomorphic coupling $g^2_h$ (there are $\ln g_h^2$ terms).
It is the physical 1PI coupling 
which controls the loop expansion and which 
will turn out to be of order $4\pi$ at the NDA scale. So, in order to use the
picture of the second of ref. \cite{nda}, it is more appropriate
to use a scheme for the Wilson effective action where the gauge coupling
is not holomorphic (see for example Ref. \cite{arkani}).
  The interesting
thing in supersymmetric theories is that the scale $\Lambda$
which determines the overall coefficient of higher dimension
operators (up to factors of order unity) can actually be determined.
This is because there is an {\it exact} result to determine
the scale, namely the gaugino condensate.

In Ref. \cite{shifman,morozov,fp}, the gaugino condensate in SUSY SU(N) was
obtained by considering first the theory with $N-1$ flavors
Higgsed far out along a flat direction.  The theory is
then weakly interacting and a reliable calculation of
the superpotential from instantons can be performed.  Once
this superpotential is obtained, the $N-1$ flavors are given a large mass,
and the gluino condensate in the low-energy SYM theory is
evaluated via the Konishi anomaly. 
The result is
\begin{equation}
\langle \lambda_a \lambda_a \rangle = 32 \pi^2 \Lambda_{DS}^3
\label{exact}
\end{equation}
where $\lambda$ is noncanonically normalized, with coefficient
of its kinetic term $1/g^2(\mu)$, and $\Lambda_{DS}$
agrees with our definition above if $g(\mu)$ is the $\drbar$ coupling.
\footnote{In Ref. \cite{fp}, the scheme independent factor $1/g^{2N}$
was never explicitly displayed; if included it would combine
together with the $g(\mu)$ in the exponent to give
the instanton amplitudes in terms of $g_h^2$.} 
 
On the other hand, according to NDA, the gaugino condensate
should be
\begin{equation}
\langle \lambda_a \lambda_a \rangle \sim \Lambda_{NDA}^3.
\end{equation}
It should be noted that NDA does not usually incorporate
large $N$ factors, but it can be easily done.
In the noncanonical normalization, the gauge propagator is 
$\sim g^2(p)/p^2$. By NDA, the condensate is going
to be saturated by loop momenta of order $\Lambda_{NDA}$, so that
\begin{eqnarray}
\langle \lambda_a\lambda_a\rangle&&\sim N^2\int{p^2dp^2\over 16\pi^2}
{g^2(p)\over p}
\sim{N^2 g^2(\lamnda)\lamnda^3\over 16\pi^2}\\ &&=N\lamnda^3
\label{loop}
\end{eqnarray}
The qualitative scaling of eq. \ref{loop} is also reproduced with the 
use of gap equations \cite{appel}.

By comparing the above to the exact result in eq. \ref{exact} one gets 
\begin{equation}
\lamnda=\left( {32 \pi^2 \over N} \right)^{1/3} \lamds.
\end{equation}
which, within an ${\cal O}(1)$ factor, coincides with $\Lambda_\infty$.
 This
is precisely what one would want to find; $\lamnda$
corresponds to the scale at which the physical coupling becomes
nonperturbative and the description of the theory must change.
Indeed, accounting also for {\it large} $N$ factors,
the loop expansion parameter of SYM should be  $x\sim Ng^2/(4\pi)^2$. Thus,
by eq. \ref{linf} we have that $\lamnda$, at which $x\sim 1$, essentially
coincides with $\Lambda_\infty$. 
In other words, we  find
out that NDA reproduces both the $4 \pi$'s and large $N$ behavior
of the instanton calculation at weak coupling. Given
that it was the physical $g(\mu)$ which appeared in the propagator,
it was natural to expect the loop counting parameter to be the 1PI
rather than Wilsonian coupling. Notice indeed that the Wilsonian
expansion parameter at the NDA scale is roughly
\begin{equation}
{Ng_W^2\over 8 \pi^2}\sim {1\over {1 +\ln (8\pi^2/N)}}
\end{equation}
which for small $N$ may look perturbative.

 As an aside, we comment on the  well known fact \cite{nsvz,rep,fp}
that the direct calculation of $\langle\lambda\lambda\rangle$ in pure, 
strongly coupled, SYM disagrees with the weak
 coupling result  calculated in the Higgsed theory. 
In particular even the  large $N$
behavior is different 
\begin{equation}
\langle \lambda\lambda\rangle_{strong}={1\over N}\langle 
\lambda\lambda\rangle_{weak}
\end{equation}
 NDA gives a result in agreement with weak
coupling methods. The origin of the discrepancy is not yet  
established. One possibility is that  while at weak coupling instantons
presumably saturate the condensate, there  are new effects, other
than  instantons, in the strong coupling regime. Quite interestingly,
due to the space time independence of the $N$-point $\lambda\lambda$
correlator, these effects should be important even at short distance
in the strong coupling regime. However, since the  scale at which
the theory became nonperturbative was adequately accounted for
in perturbation theory, both instantons and these additional
nonperturbative effects seem irrelevant to establishing the NDA scale.
Finally,
another possibility, recently proposed by Kovner and Shifman, is that there
exists a chirally symmetric phase of SYM \cite{kovner}. In this way the 
strong coupling result, interpreted as a weighted average over
chirally symmetric and asymmetric vacua, is bound to be the lower.

The obvious question now is how to extrapolate these lessons
to real QCD.  
In QCD, we know that a 2 loop calculation, although
it goes in the right direction, does not
change the QCD scale sufficiently to account for NDA.
In fact,the relation between scales in this case
is given 
\begin{equation}
\Lambda_{2-loop}\simeq \left ({3.2\pi^2\over g^2}\right )^{0.33} 
\Lambda_{1-loop}
\end{equation}
where 5 flavors have been assumed (but the result is rather
independent on that). By running from $\mu=M_Z$,
the above equation changes the one-loop QCD scale of about 100 MeV to its
two-loop value of about 250 MeV. This is still about 4 times smaller 
than the ``observed'' NDA scale of QCD. As we said before,
higher loop effects, or scheme dependence, in the perturbative
definition of $\lamnda$ could amount to an ${\cal O}(1)$ factor.
So the ``big'' factor in QCD is puzzling. In the next section we
will argue that additional corrections could come from instantons.
This gives rise to the obvious question of why the
instanton effects do not affect the ``exact'' coupling
of $N=1$ SYM.  We will address this issue in the
concluding section. The discussion of this section seems to suggest that additional 
effects,
if present, are not very important in determining the strong scale,
as the condensate is well estimated perturbatively.


Finally we briefly comment on the relation, suggested by NDA, between
the hadron masses and their sizes, as determined by
the confinement scale. Consider
for instance the elastic scattering of a spin zero glueball $S$ of mass
$\sim \Lambda_{NDA}$. By NDA we 
expect the quartic coupling to be ${\cal O}(16\pi^2)$, so that the elastic cross
section is $\sim 4 \pi (4 \pi/\lamnda)^2$. This result can be interpreted
as due to collision of two hard objects of radius $4 \pi/\lamnda$.
This remains true also at large $N$ though there the mesons are weakly
interacting, and the suitable factor of $N$ must be factored out in the 
cross section. It should be noted that from this point of view, the coincidence
of $\Lambda_{QCD}$, the scale at which the perturbative gauge coupling
blows up, and the confinement scale, $250 {\rm MeV} \approx 2 \lamnda/4 \pi$, 
is merely coincidental.
The first scale seems to have no physical meaning, though it can of course
be defined from the two-loop $\beta$-function.

What we have learned about NDA in SYM can be useful in particle physics
models where the strong interaction scale $\Lambda$ itself depends on some
modulus $X$. In the models of refs. \cite{plateau,infla,luty} $X$ gives
 mass to 
all $SU(N)$ flavors, so that below the scale $X$ the effective gauge theory 
is just SYM. The strong dynamics then generates a superpotential
for $X$ via gaugino condensation. In refs. \cite{plateau,infla} the effective
superpotential is $W_{eff}=\Lambda^3_{DS}(X)=\Lambda^2 X$, 
where $\Lambda_{DS}(X)$ and $\Lambda$ are respectively the holomorphic scales
in the low- and high-energy theories.
 The resulting potential is very flat, and an estimate
of the non-perturbative Kahler potential can be crucial. 
The only way we can do that at the
 moment is by  NDA. In ref. \cite{infla} $X$ is the inflaton so that
the size and sign of these corrections can have a crucial impact on the slow
roll.  We have argued that NDA  should apply for SYM.
Denoting by $\Lambda_{NDA}(X)$ and $\Lambda_{DS}(X)$ the scales of
the low energy theory the effective lagrangian is \cite{luty}
\begin{eqnarray}
{\cal L}_{eff}&=&{1 \over 16 \pi^2}(\int d^4\theta 
\,c_1\,\lamnda(X)^\dagger\lamnda(X)\\
&+&\int d^2\theta\, \lamnda(X)^3 +{\rm h.c.})
\label{leff}
\end{eqnarray}
where $c_1$ is expected to be $O(1)$. By writing the above in terms of 
the original holomorphic $\Lambda$ we have that the correction 
to the Kahler metric is
\begin{equation}
\delta K_{XX^\dagger}= {c_1\over 9} \left ({1\over 4 \pi}\right )^{2/3}
\left ({\Lambda\over X}\right )^{4/3}.
\label{metric}
\end{equation} 
This scaling with $4 \pi$ can also be established by a direct diagrammatic
analysis. For this purpose it is useful to parameterize $X=\langle X\rangle
+\delta X$, where $\langle X\rangle$ is the c-number VEV.
Below the scale $\langle X\rangle$, where the messengers are integrated out,
$\delta X$ couples effectively to the SYM theory via a one-loop effect
\begin{equation}
{1\over 8\pi^2}\int d^2\theta {\delta X\over \langle X\rangle}W_\alpha
W^\alpha +\dots.
\label{shrink}
\end{equation}
At second order in this interaction, we get a quantum correction to the
$\delta X$ wave function
\begin{equation}
\delta K_{XX^\dagger}=\left ({1\over 8\pi^2}\right )^3\int_{0}^{
\langle X\rangle^2} dp^2 {g^4(p^2)\over 
\langle X\rangle^2}
\label{direct}
\end{equation}
where two powers of $1/8\pi^2$ come from insertions of
eq. \ref{shrink}. For large $\langle X\rangle$
the leading contribution to eq. \ref{direct} is 
perturbative $\sim g^4(\langle X\rangle)/(4\pi)^6$ and comes from 
integration at  $p\sim \langle X\rangle$.
 Non-perturbative effects are estimated via
NDA by the contribution at $p\sim \Lambda_{NDA}$ to eq. \ref{direct}.
The result agrees with the estimate in eq. \ref{metric}. This effect becomes
rapidly important when $\langle X\rangle$ is decreased.

It would be interesting to have information on the sign of this
correction. Notice that $\delta K_{XX^\dagger}$
is proportional to the correlator $G_{\lambda}=\langle (\lambda\lambda )
(\bar\lambda\bar\lambda)\rangle$ at zero momentum\footnote{
$\delta K_{XX^\dagger}$ can be obtained by focusing on the correction
to $F_XF_X^\dagger$ and by noticing that $F_X$ couples just 
to $\lambda\lambda$ in eq. \ref{shrink}.}
It is useful to consider the dispersive
K\"allen-Lehman representation for $G_\lambda$
\begin{equation}
G_\lambda(p^2)=\int_0^\infty dm^2 {m^2 \sigma(m^2)\over p^2-m^2}
\label{kallen}
\end{equation}
where $\sigma(m^2)$ is a positive definite function. \footnote{
 Notice that $\sigma\sim g^4(m^2)$, at $m\gg \Lambda_{eff}$
giving an apparent quadratic divergence.
This is regulated by performing the suitable matching to the above theory
at the scale $\langle X\rangle$. Above this scale there is just
the logarithmic divergence of the $X$ wave function}
Our basic point is that we can perform an OPE for $G_\lambda$
in the Euclidean $-p^2>>\Lambda_{eff}^2$ region. The lowest power correction
corresponds to $\langle GG\rangle/p^4\sim \Lambda_{eff}^4/p^4$.
It is reasonable to expect that the corrections to 
$\sigma(m^2\gg \Lambda_{eff}^2)$ scale in the same way. 
If that is the case, the leading non-perturbative contribution
to $\delta K_{XX^\dagger}\propto G_\lambda(p^2=0)$ comes  only 
from the region of integration at $m^2\sim \Lambda_{eff}^2$. This 
gives more justification to the NDA estimate we did above.
This also suggests that it is probably not unreasonable
to assume that  the effect is well described
by summing just over the lowest resonances in $\sigma$.
Under that assumption 
we conclude that $c_1>0$ in eq. \ref{metric}. This result has 
important implications for the models of 
refs \cite{plateau,infla}. There  the correction with $c_1>0$ 
creates a potential that ``pushes'' the $X$ field towards the origin.
This is not problematic for the inflationary models of ref. \cite{infla},
but it can destabilize the local vacuum of the gauge mediated scenarios
of ref. \cite{plateau}. The condition to avoid the latter problem has
already been discussed in ref. \cite{plateau}: it requires the scale $X$
to be above $10^{9}-10^{10}$ GeV.

\section{$N=2$ SUSY SU(2)}

For N=2 SUSY QCD the effective Lagrangian
to leading order in a momentum expansion
 was derived in an exact solution by Seiberg and Witten \cite{SW}. Although the dynamics and physical
fields are very different from QCD, one can nonetheless
observe a similar puzzle; 
the  effective coupling blows up at the point where
according to the one-loop beta function one would get
 $g^2/(4\pi)=\alpha=0.76$.  This might have been thought to be
a safely perturbative region, but in fact  
the instanton effects become  
 large at this point, they  induce very strong interaction
and make the use of the original formulation  of the theory impossible.

This is seen from the exact result for the effective coupling 
\begin{equation}
{8\pi \over g^2(u)} = {K(\sqrt{1-k^2}) \over K(k)}
\end{equation}
where K is elliptic integral and the argument
\begin{equation}
k^2={(u-\sqrt(u^2-4 \Lambda^4))\over (u + \sqrt{u^2-4 \Lambda^4}))}
\end{equation}
is a function of gauge invariant vacuum expectation of squared scalar field 
\begin{equation}
u={1\over 2}<\phi^2>={a^2 \over 2} +{\Lambda^4 \over a^2} + \ldots
\end{equation}
and a is just its VEV. For large $a$ there is a weak coupling expansion
which includes  instanton effects  \footnote{It should
be noted that the first terms in this expansion have
been explicitly verified in instanton calculations \cite{inst}.}
\begin{equation} \label{pert}
{8\pi \over g^2(u)}={2 \over \pi} \left ( \log \left( {2 a^2 \over \Lambda^2 }\right) -
 {3 \Lambda^4 \over a^4}+ \ldots \right)
\end{equation}
The exact coupling blows up at $u=2 \Lambda^2$, which means that the factor
between the exact strong interaction scale and the perturbative
one is in this theory $\Lambda_{\infty}=2^{3/2}\lamds$. Actually
this is the ratio of the scale $\sqrt{u}$ to the scale $a$
at which the perturbatively evolved coupling (one-loop) blows up.
If one were to account for the next term in the expansion of $u$,
the ratio of scales is reduced to $\sqrt{2} \sqrt{2+\sqrt{2}}$.
The fact that instanton effects can be important at such a high
scale was anticipated in Ref. \cite{cdg} and is presumably due
to the significance of the prefactor in instanton calculations.)

The behavior is shown in  Fig.(\ref{figure1}), where we
have included both a curve which shows the full 
coupling (thick solid line), as well as a curve which illustrates
only the one-instanton correction (thick dashed one). Because we will want
to compare the running of the coupling in different theories,
we have plotted $b g^2/8 \pi^2$ (b=4 in this case is the one-loop coefficient
of the beta function) and measure all quantities 
in  units of $\lamds$, so that  the one-loop
charge blows out at 1. The meaning of the scale
can therefore be determined by what enters in the logarithm. We have
plotted the exact result against $\phi$ rather than $a$.

Note the very rapid change of the coupling induced by instantons.
It is also of interest that the full multi-instanton sum makes
the rise in the coupling even more radical than with only the
one-instanton correction incorporated. It is also interesting
to observe that at the scale where the true coupling blows up,
the perturbatively evolved coupling is still not very large.
 Individually, the perturbative log and instanton corrections
are well defined at this region: however they cancel each other in the
inverse charge. This
is encouraging from the point of view of developing a consistent
expansion for the
instanton corrections. The rapid rise in the coupling is also
encouraging in that it ensures that perturbation theory is valid
almost to the point where it blows up. For a consistent
picture of QCD, in which perturbation theory still appears to
be applicable at the $c$-quark scale, while the theory
is nonperturbative at 1 
GeV, such a dramatic effect is essential.

So the basic lessons seems to be first, that instantons can
lead to a discrepancy between the scales at which the perturbative
evolution of the coupling blows up and the nonperturbative coupling blows up.
Second, the rise in the coupling is very dramatic.  Third, the coincidence
between $f_\pi$ and $\Lambda_{QCD}$ seems to be just that. Fourth,
even though we are in a weak coupling regime where the instantons
should be dilute, there is a notable difference between the rise
in the coupling due to one instanton and due to the full instanton sum;
multi-instantons are also important.

\section{QCD}
Having learned what we can from exact SUSY results, we
now turn to  ordinary QCD. We know that naive dimensional
analysis (NDA) appears to work, but with the dimensional
scale which sets the cutoff and which suppresses higher
dimensional operators set by 1 GeV. As the theory
is not supersymmetric, we cannot derive this dimensional
scale as we did with the gluino condensate. Nonetheless,
sum rules \cite{sumrule} and  lattice simulations \cite{gupta} yield, 
consistently,
a quark condensate $\langle \bar q q\rangle\sim (250\,{\rm MeV})^3$.
On the other hand, NDA suggests
 \begin{equation}
 \langle \bar q q \rangle= {\lamnda^3\over 16 \pi^2 Z(\lamnda)}
\sim {(1\, {\rm GeV})^3\over 16 \pi^2 Z(\lamnda)}
\label{condensate1}
\end{equation}
where $Z$ is the multiplicative renormalization of the composite 
operator $\bar q q$. Eq. \ref{condensate1}  agrees with the
phenomenological value of the condensate if $Z(\lamnda)\sim 1$.
This requirement is not inconsistent as $Z$ does not run very fast.

 The question then is where this scale arises in terms of the
theory of the fundamental fields, the quarks and gluons.  We know
that the parton model is limited from below by the 1 GeV scale,
but   perturbative QCD seems
well-behaved at this scale.
For the $N=1$ coupling, we know we can estimate the dimensional
analysis scale using only the leading two terms in perturbation
theory. We also know this is inadequate in QCD, as the subleading
term changes the QCD scale from about 100 MeV to 250 MeV;
higher order terms should not change this substantially.

So it seems that if the NDA scale of 1 GeV is to have a physical
meaning it is probably due to nonperturbative effects
in the evolution of the coupling. It is possible that
instantons are the only nonperturbative effect which is
substantial above the confinement scale. This was the original
CDG suggestion \cite{cdg} . For a review of subsequent
developments and phenomenological
fits in support of this hypothesis 
with allowance for inter-instanton interactions
(the ``instanton liquid model'')  one may consult the review \cite{SS_97}.

  For the purpose of this discussion we restrict ourselves to  the one
 instanton effect (though
for QCD the distribution allows for instanton interactions), a single-instanton correction
 to the effective action of some smooth background gauge
field $G_{\mu\nu}$. 
 The external field is supposed to be normalized at some normalization scale
$\mu$, and CDG has proposed to include all instantons with size $\rho
<\rho_{max}=1/\mu$. The effective charge is then defined as:
\begin{equation} \label{eq_CDG}
{8\pi^2 \over g^2_{eff}(\mu)} = b \, \ln({ \mu\over \Lambda_{pert}}) 
\\ \nonumber
- {4\pi^2 \over (N_c^2-1)} \int^{\rho_{max}}_0 dn(\rho) \rho^4 
({8\pi^2 \over g^2_{eff}(\rho)})^2 
\end{equation}
  where 
$b=11N_c/3-2N_f/3$ is the usual one-loop coefficient of the beta function, and $dn(\rho)$ is the distribution of instantons 
(and anti-instantons) over  size.

 The instanton
density is
semiclassically calculable only at small $\rho$ where it
 is very small: 
and therefore for high scales  
 the instanton correction is tiny, $\sim (\Lambda/\mu)^b$.
 At larger $\rho$ the instanton size distribution
 (known  from lattice studies and model-dependent calculations)
has a strongly  peaked shape, with the peak at $\rho\sim .2-.3 {\rm fm}$ 
\cite{Shu_twoscales}. 
 As soon as $\rho_{max}=1/\mu$ becomes close to 
  the position of the peak, 
  the CDG effective charge (\ref{eq_CDG}) blows up. 

Of course in order to establish the magnitude of the effect
of instantons one needs to know the instanton distribution.
In Figure (\ref{figure1}) we compare three QCD-like theories
The first two are pure gauge SU(2) and SU(3) where
the distributions are obtained by cooling of lattice configurations
from \cite{SU2_lat} and \cite{SU3_lat} respectively {There are
 other recent lattice works for SU(2) and SU(3) which we have 
not included, e.g.
\cite{SU2_lat2} and \cite{SU3_lat2}}. For QCD with three light
flavors, we use the Interacting Instanton Liquid Model (IILM) \cite{Shu_sizes}
to provide a model distribution.

The two-loop result (shown by  the thin short-dashed line) is the same for
any pure gauge theory (with minor modifications for QCD  which are not shown).
 As we have mentioned, the two-loop running raises
the scale at which the coupling gets big,
but  does not generate a sufficiently large scale.
With instantons present, the coupling blows up at a somewhat
higher scale, where perturbation theory looks naively still valid.   
It is perhaps a little surprising that all the theories including
instantons look so similar, and that furthermore the instanton
effects are so similar to that for N=2 SUSY QCD. Although
this is probably accidental, it does establish that
instantons can generate a somewhat higher scale, and that
the theory looks perturbative until almost reaching this scale
(that is the coupling rises very rapidly). 
One may further speculate by analogy to the N=2 theory that
multi-instanton effects  yield an even
more substantial rise in the coupling.

\begin{figure}[t]
\vskip -0.4in
\epsfxsize=3.8in
\centerline{\epsffile{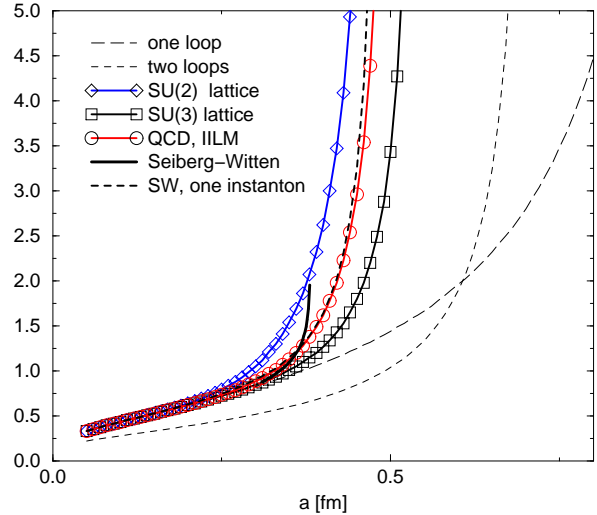}}
\vskip -0.05in
\caption[]{
 \label{figure1}
 The effective charge $b \,g^2_{eff}(\mu)/8\pi^2$ (b is the coefficient
of the one-loop beta function) versus normalization scale $\mu$ (in units of
its value at which the one-loop charge blows up). The thick solid line
correspond to exact solution \cite{SW} for the N=2 SYM, the thick dashed line
shows the one-instanton correction. Lines with simbols (as indicated on figure)
stand for N=0 QCD-like theories,
SU(2) and SU(3) pure gauge ones and QCD itself. Thin long-dashed and short-dashed lines are one and tho-loop results.
}
\end{figure}

\section{Discussion and Conclusions}

In conclusion, we have argued that in $N=1$ pure gauge theory
the analog of the ``chiral lagrangian'' scale
corresponds to the scale where
the physical (1PI) coupling becomes of order $4 \pi$. 
We had two benefits from considering the supersymmetric theory;
the NDA scale is {\it determined} by the gluino condensate and
furthermore the exact beta function is known.
Our argument is made by comparing the known exact value
of the gluino condensate to a ``naive dimensional analysis estimate'',
and by evaluating the exact coupling at this scale.
This is a useful conclusion for applications to supersymmetric model
building \cite{nda,kaplan,nelstra,plateau,infla,luty}. The lessons for QCD 
are   less obvious. In QCD the chiral lagrangian scale $\Lambda_{NDA}$
is in fact somewhat larger ($\sim 1$ GeV) than the 
scale $\Lambda_{QCD}\sim 250$ MeV which is associated to the
perturbative coupling. It was pointed out almost two decades ago \cite{cdg},
that  small instantons  can lead 
to a precocious breakdown of perturbation theory, {\it i.e.} at scale
where $\alpha/4\pi$ is somewhat less than unity. 
It is an important question how instanton effects
can be relevant for QCD but not relevent for $N=1$ supersymmetric
theories. 
 
Consider, for example, $N=0$ SU(3) with three flavors
and $N=1$ SU(3) with no flavors. These theories have
identical $b_0$ and numerically very similar $b_1$.  Since
the perturbative scaling is therefore practically the same,
we need to understand whether it is possible that
instanton effects can be important for the first theory
at a scale in which they can be neglected in the second.

To address this question, we first need to remind ourselves
how instantons can affect the coupling.  The naive answer is
that they don't since the exact beta function is given
without including instanton corrections to the running. In
essence supersymmetry protects against contributions
from small instantons. However, this answer is
inadequate since we already know  
  that the dominant instanton  contribution
will only affect the coefficient
of 
a multifermion operator. This only contributes to the renormalization
of the gauge coupling in the presence of a nonzero gluino condensate,
which is only relevant in the infrared and would not be included
in the definition of the beta function.

Now let us compare our two theories in more detail. We need to
determine the normalization by which we will compare the theories.
This is straightforward as we can take identical gauge coupling
values in the ultraviolet for the two theories in a safely perturbative
regime.  Subsequently, running down in energy, both couplings
will run according to perturbation theory essentially identically.
Actually there is a caveat that the coupling we generally use
in QCD is the $\overline{MS}$ coupling whereas the $\overline{DR}$
scheme is used in the $N=1$ theory. However, one
can explictly check that the difference in couplings
is sufficiently small that it does not affect the argument below.

Now for the $N=1$ theory we know the NDA scale by the arguments
given previously. In fact we know it is the scale at which
the coupling blows up, that is about 250 MeV.  

Let us consider the fermion condensates and their evolution in the
two theories (we mean here the condensates of the fields that are canonically
normalized at each scale). At the scale where $g^2\sim 1$ the two
condensate are roughly the same since
$\langle \lambda\lambda\rangle=(\lamnda^{N=1})^3=(250 {\rm MeV})^3$ and
$\langle \bar q q\rangle= (1 {\rm GeV})^3/16 \pi^2\simeq (250 {\rm MeV})^3$.
Now let us compare their values at 1 GeV. The gluino condensate scales
inversely to the gauge coupling squared and is therefore 
$(250 {\rm MeV})^3/g^2(1 {\rm GeV})\simeq (250 {\rm MeV})^3/4\pi$.  
This is smaller
than the quark condensate which is scaling with the mass anomalous
dimension, that is, much more slowly, and is therefore still about
$(250 {\rm MeV})^3$. That is, the QCD condensate is
significantly larger. It is therefore not inconsistent
for instantons to be significant at 1 GeV in QCD, but not
in the comparable supersymmetric theory. Although this
does not establish the importance of instantons for QCD, it is reassuring.

In order to probe the possible relevance of instanton effects, we studied
the effective gauge coupling in the $N=2$ $SU(2)$ SYM.
There also we found that instanton effects lead to a precocious
explosion of the coupling. We compared the effective charge
as a function of the adjoint VEV in N=2 with the effective charge
of $N=0$ gauge thories obtained by smearing
over small instanton effects. We used the 
instanton liquid model and lattice data to estimate the instanton
density. We found a suggestive similarity (see Fig. 1) between these
phenomenological models and the exact $N=2$ case.
Furthermore, the generated scales are in the same ballpark, and
for QCD it is not so far from 1 GeV, the right phenomenological
value.   
All this seem to support the instanton scenario of the generation of
this scale for QCD, although it does not exclude other nonperturbative
effects.

In summary, we have looked at potential implications of SUSY theories
for our understanding of low energy QCD. Because the phase structure of
supersymmetric
theories is so different, we restricted attention to the question
of obtaining a consistent picture of
the boundary of the perturbative domain.
We have not addresssed the issue of
chiral symmetry breaking and confinement (aside from
mention of the geometric relation between the confinement scale
and the chiral lagrangian scale).
We have also not addressed
the possible discrepancies in the effective coupling which
can appear depending on the correlation function at low energy.
We have argued that in fact the fundamental QCD scale of the
theory is more readily that associated with the blow-up of the physical
coupling, and that this is more likely to be the higher chiral lagrangian
scale. We also note that the separation of scales due to instantons
argues
against a standard large-$N$ interpretation of QCD,  for which
instantons would not be important.

We acknowledge useful conversations with N. Arkani-Hamed,
M. Beneke, C. Callan, A. Cohen, D. Finnell, H. Georgi,
 R. Jaffe, K. Johnson, D. Kaplan, M. Luty, P. Nason, J. Negele, A. Nelson,
M. Porrati, M. Strassler and F. Wilczek. L. R. thanks Princeton University and
the Institute for Advanced Study for their hospitality during
the course of this work.

\end{narrowtext}

\end{document}